\documentclass[reprint, amsmath,amssymb, aps, floatfix]{revtex4-2}

\usepackage{graphicx}
\usepackage{dcolumn}
\usepackage{bm}
\usepackage[colorlinks = true,linkcolor = blue,urlcolor  = blue,citecolor = blue,anchorcolor = blue]{hyperref}
\usepackage{txfonts}

\usepackage{microtype}

\begin{document}

\preprint{APS/123-QED}

\title{Thermal effects on tidal deformability of a coalescing binary neutron star system}

\author{A. Kanakis Pegios}
\author{P.S. Koliogiannis}%
 \email{pkoliogi@physics.auth.gr}
\author{Ch.C. Moustakidis}
\affiliation{Department of Theoretical Physics, Aristotle University of Thessaloniki, 54124 Thessaloniki, Greece}

\date{\today}

\begin{abstract}
The study of neutron star mergers by the detection of the emitted gravitational waves is one of the most promised tools to study the properties of dense nuclear matter at high densities. It is worth claiming that, at the moment, strong evidence that the temperature of the stars is zero during the last orbits before coalescing does not exist. Contrariwise, there are some theoretical predictions suggesting that the star's temperature might even be a few MeV. According to the main theory, the tides transfer mechanical energy and angular momentum to the star at the expense of the orbit, where friction within the star converts the mechanical energy into heat. During the inspiral these effects are potentially detectable. Different treatments have been used to estimate the transfer of the mechanical energy and the size of the tidal friction, leading to different conclusions about the importance of pre-merger tidal effects. The present work is dedicated to the study of the effect of temperature on the tidal deformability of neutron stars during the inspiral of a neutron star system just before the merger. We applied a class of hot equations of state originated from various nuclear models and found that even for low values of temperature ($T<1$ MeV) the effects on the basic ingredients of tidal deformability are not negligible. However, according to the main finding, the effect of the temperature on the tidal deformability is indistinguishable. The consequences of this unexpected result are discussed and analyzed.
\end{abstract}


\maketitle

\section{Introduction}
The last years very rich information has been gained by the detection of gravitational waves of binary black holes, binary neutron stars, and black hole-neutron star systems by the LIGO/Virgo collaboration~\cite{Abbott-gw170817,Abbott-gw190425}. In particular, the binary neutron stars system during the inspiral, as well the merger and post-merger epoch, is considered as one of the best extraterrestrial laboratories to study unexplored properties of dense cold and hot nuclear matter, mainly through the tidal deformability~\cite{Kanakis-2020,Kanakis-2021,Margaritis-2021}.   

It is worth claiming that strong evidence that the temperature of the stars is zero during the last orbits before coalescing does not exist. Contrariwise, there are some theoretical statements or conjectures that the temperature of the star might even be a few MeV~\cite{Meszaros-92,Bildsten-92,Kochanek-92}. Following the very recent study of Arras and Weinberg~\cite{Arras-2019} tides transfer mechanical energy and angular momentum to the star at the expense of the orbit, and then friction within the star converts the mechanical energy into heat. During the inspiral, these effects are potentially detectable as a deviation of the orbital decay rate from the general relativistic point-mass result, or as an electromagnetic precursor if heating ejects the outer layers of the star. Different treatments have been used to estimate the transfer of energy and the size of the tidal friction, leading to different conclusions about the importance of the pre-merger tidal effects~\cite{Arras-2019}.

Firstly, Meszaros and Rees~\cite{Meszaros-92} stated that during the final spiraling of the binary neutron star system, the two stars will distort each other tidally. These effects lead to an appreciable heat before the merger. It is stated that due to the effects of tidal heating, the neutron star will heat up to a temperature (both core and crust) comparable to those of supernova explosions~\cite{Meszaros-92}. Similar results have been found in Refs~\cite{Bildsten-92,Kochanek-92}. Other studies concerning the tidal heating, due to resonant excitation of g-mode, inertial moment resonances, bulk and shear viscosity, and superfluid effects, have been presented in Refs.~\cite{Lai-94,Reisenegger-94,Ho-99,Lai-2006,Xu-2017,Yu-2017}. Recently, Arras and Weinberg~\cite{Arras-2019} studied the impact of Urca reactions driven by tidally induced fluid motion during binary neutron star inspiral~\cite{Arras-2019}. 

All the above studies conclude that tidal heating effects, during the inspiral, are present (due to different reasons) and also lead to the heating of the interior of neutron stars. The predictions cover a large interval, higher than three orders of magnitude, from $T=0.01$ MeV to $T=10$ MeV. The tidal deformability is a quantity sensitive to the equation of state of neutron stars. In particular, it depends both on the tidal Love number $k_2$ and the radius $R$. Since both quantities are determined within the applied equation of state and for a fixed value of the mass, it will be of great interest to study how thermal effects deviate both quantities from the cold case and mainly to what extent. This is important since the tidal deformability, and consequently the corresponding radius, are constrained from the detection of the gravitational waves.

The motivation of the present work is to examine to what extent the temperature, due to various mechanisms, affects the values of the tidal deformability of a neutron star during the inspiral process, just before the merger. According to our knowledge, this possibility has never been examined. We concentrate on the case of isothermal matter in the interior of the neutron stars, employing some of the most used equations of state, for temperatures in the range of $T=0.01-1$ MeV~\cite{Lattimer-91,Shen-2011,Steiner-2013,Banik-2014,Koliogiannis-2021}. Moreover, for reasons of completeness and comparison, we extend the study to the more extreme but less reliable case of even higher temperature.

\section{Tidal deformability} \label{sec:tidal}
The emitted gravitational waves from the late phase of the inspiral, before the merger, are a very important source for the detectors~\cite{Postnikov-2010,Flanagan-08,Hinderer-08}, leading to the measurement of various properties. During this phase, the tidal effects can be detected~\cite{Flanagan-08}.

The tidal Love number $k_2$, which depends on the equation of state, describes the response of a neutron star to the presence of the tidal field $E_{ij}$~\cite{Flanagan-08}. This relation is given below
\begin{equation}
Q_{ij}=-\frac{2}{3}k_2\frac{R^5}{G}E_{ij}\equiv- \lambda E_{ij},
\label{Love-1}
\end{equation}
where $R$ is the neutron star radius and $\lambda=2R^5k_2/3G$ is the tidal deformability. The tidal Love number $k_2$ is given by \cite{Flanagan-08,Hinderer-08}
\begin{eqnarray}
k_2&=&\frac{8\beta^5}{5}\left(1-2\beta\right)^2\left[2-y_R+(y_R-1)2\beta \right]\nonumber\\
& \times&
\left[\frac{}{} 2\beta \left(6  -3y_R+3\beta (5y_R-8)\right) \right. \nonumber \\
&+& 4\beta^3 \left.  \left(13-11y_R+\beta(3y_R-2)+2\beta^2(1+y_R)\right)\frac{}{} \right.\nonumber \\
&+& \left. 3\left(1-2\beta \right)^2\left[2-y_R+2\beta(y_R-1)\right] {\rm ln}\left(1-2\beta\right)\right]^{-1},
\label{k2-def}
\end{eqnarray}
where $\beta=GM/Rc^2$ is the compactness of a neutron star. The quantity $y_R$ is determined by solving the following differential equation
\begin{equation}
r\frac{dy(r)}{dr}+y^2(r)+y(r)F(r)+r^2Q(r)=0, 
\label{D-y-1}
\end{equation}
with the initial condition $ y(0)=2$~\cite{Hinderer-10}. $F(r)$ and $Q(r)$ are functionals of the energy density ${\cal E}(r)$, pressure $P(r)$, and mass $M(r)$ defined as~\cite{Postnikov-2010}
\begin{equation}
F(r)=\left[ 1- \frac{4\pi r^2 G}{c^4}\left({\cal E} (r)-P(r) \right)\right]\left(1-\frac{2M(r)G}{rc^2}  \right)^{-1},
\label{Fr-1}
\end{equation}
and
\begin{eqnarray}
r^2Q(r)&=&\frac{4\pi r^2 G}{c^4} \left[5{\cal E} (r)+9P(r)+\frac{{\cal E} (r)+P(r)}{\partial P(r)/\partial{\cal E} (r)}\right] \nonumber \\
&\times&
\left(1-\frac{2M(r)G}{rc^2}  \right)^{-1}- 6\left(1-\frac{2M(r)G}{rc^2}  \right)^{-1} \nonumber \\
&-&\left[\frac{4M^2(r)G^2}{r^2c^4}\left(1+\frac{4\pi r^3 P(r)}{M(r)c^2}   \right)^2\right. \nonumber \\ &\times&\left. \left(1-\frac{2M(r)G}{rc^2}  \right)^{-2}\right].
\label{Qr-1}
\end{eqnarray}
Eq.~(\ref{D-y-1}) must be solved numerically and self consistently with the Tolman - Oppenheimer - Volkoff (TOV) equations using the boundary conditions $y(0)=2$, $P(0)=P_c$ ($P_{c}$ denotes the central pressure), and $M(0)=0$~\cite{Postnikov-2010,Hinderer-08}. From the numerical solution of TOV equations, the mass $M$ and radius $R$ of the neutron star can be extracted, while the corresponding solution of the differential Eq.~(\ref{D-y-1}) provides the value of $y_R=y(R)$. This parameter along with the quantity $\beta$ are the  basic ingredients  of the tidal Love number $k_2$.

The chirp mass {\it $\mathcal{M}_c$} of a binary neutron stars system is a well measured quantity by the detectors~\cite{Abbott-gw170817}. Its exact form is given below
\begin{equation}
\mathcal{M}_c=\frac{(m_1m_2)^{3/5}}{(m_1+m_2)^{1/5}}=m_1\frac{q^{3/5}}{(1+q)^{1/5}},
\label{chirpmass}
\end{equation}
where $m_1$ is the mass of the heavier component star and $m_2$ is the lighter's one. Therefore, the binary mass ratio $q=m_2/m_1$ lies within the range $0\leq q\leq1$.

Also, another quantity that is well measured is the effective tidal deformability $\tilde{\Lambda}$ which is given by~\cite{Abbott-gw170817}
\begin{equation}
\tilde{\Lambda}=\frac{16}{13}\frac{(12q+1)\Lambda_1+(12+q)q^4\Lambda_2}{(1+q)^5},
\label{L-tild-1}
\end{equation}
where $\Lambda_i$ is the dimensionless deformability~\cite{Abbott-gw170817}
\begin{equation}
\Lambda_i=\frac{2}{3}k_2\left(\frac{R_i c^2}{M_i G}  \right)^5\equiv\frac{2}{3}k_2 \beta_i^{-5}  , \quad i=1,2.
\label{Lamb-1}
\end{equation}
The effective tidal deformability $\tilde{\Lambda}$ plays important role on the neutron star merger process and is one of the main quantities that can be inferred by the detection of the corresponding gravitation waves.
\section{Results and Discussion}
In order to study the thermal  effects on the tidal deformability, we employ a class of hot equations of state derived in the past. All of them have been applied successfully in the study of bulk properties of  hot neutron stars, proto-neutron stars, and supernovae. It is worth pointing out that thermal effects are more pronounced in the case of low neutron star mass, which corresponds to the region of crust, while the effects on the core matter, at least up to 1 MeV, are negligible. We expect that this particular effect, due to quantum behavior of matter, may be reflected not only on the mass-radius diagram of neutron stars for a specific hot equation of state, but also on some specific properties that are sensitive on the structure of neutron star, such as the tidal deformability. In particular, we first used the hot equations of state of Lattimer and Swesty~\cite{Lattimer-91}, Shen \textit{et al.}~\cite{Shen-2011}, Steiner \textit{et al.}~\cite{Steiner-2013}, and Banik \textit{et al.}~\cite{Banik-2014}. Moreover, we also employed the MDI-APR equation of state~\cite{Akmal-1998}. The corresponding model in the latter case has been applied successfully in similar studies and it is suitable for the description of both cold and hot neutron star matter~\cite{Koliogiannis-2020,Koliogiannis-2021}.

\begin{figure*}
\includegraphics[width=0.49\textwidth]{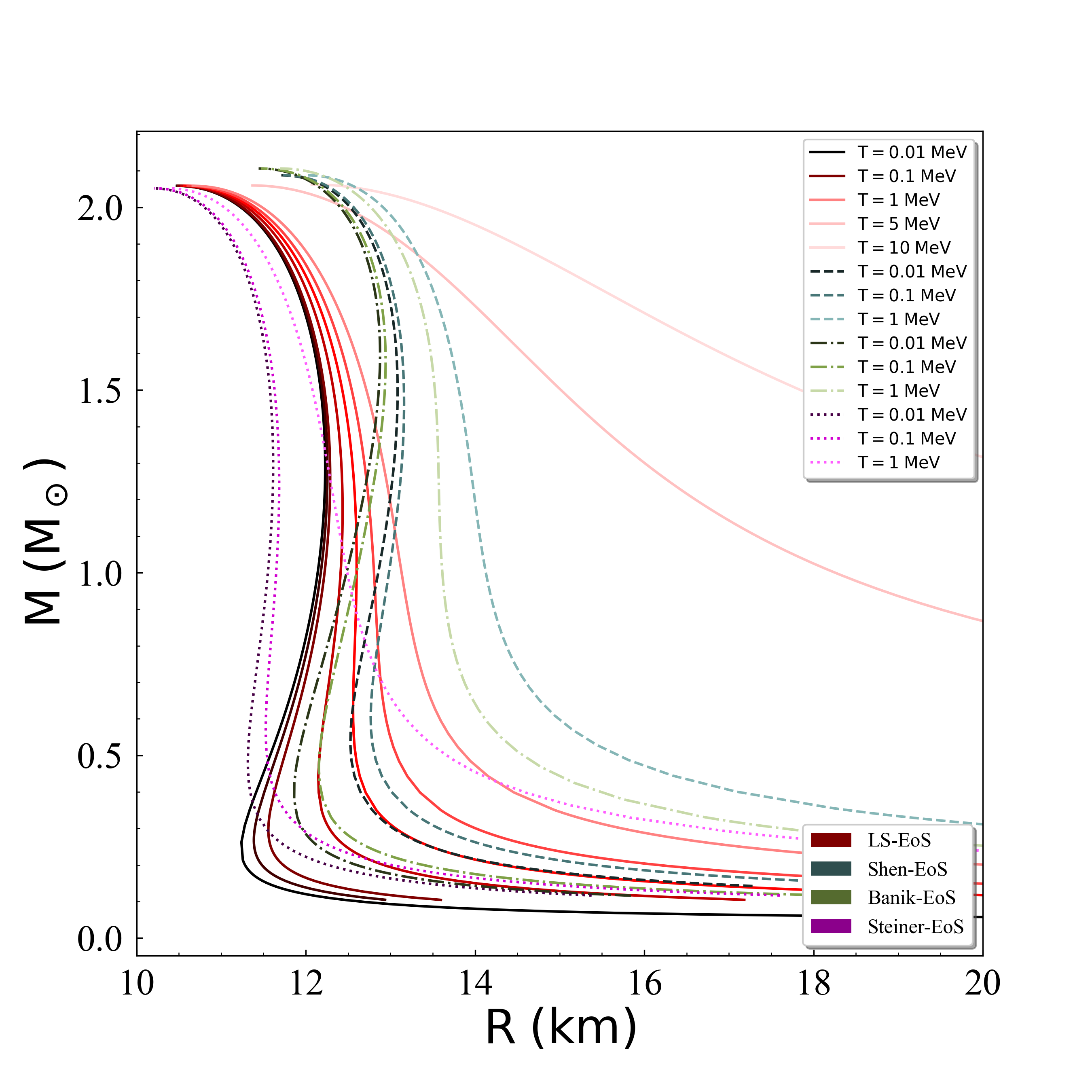}
~
\includegraphics[width=0.49\textwidth]{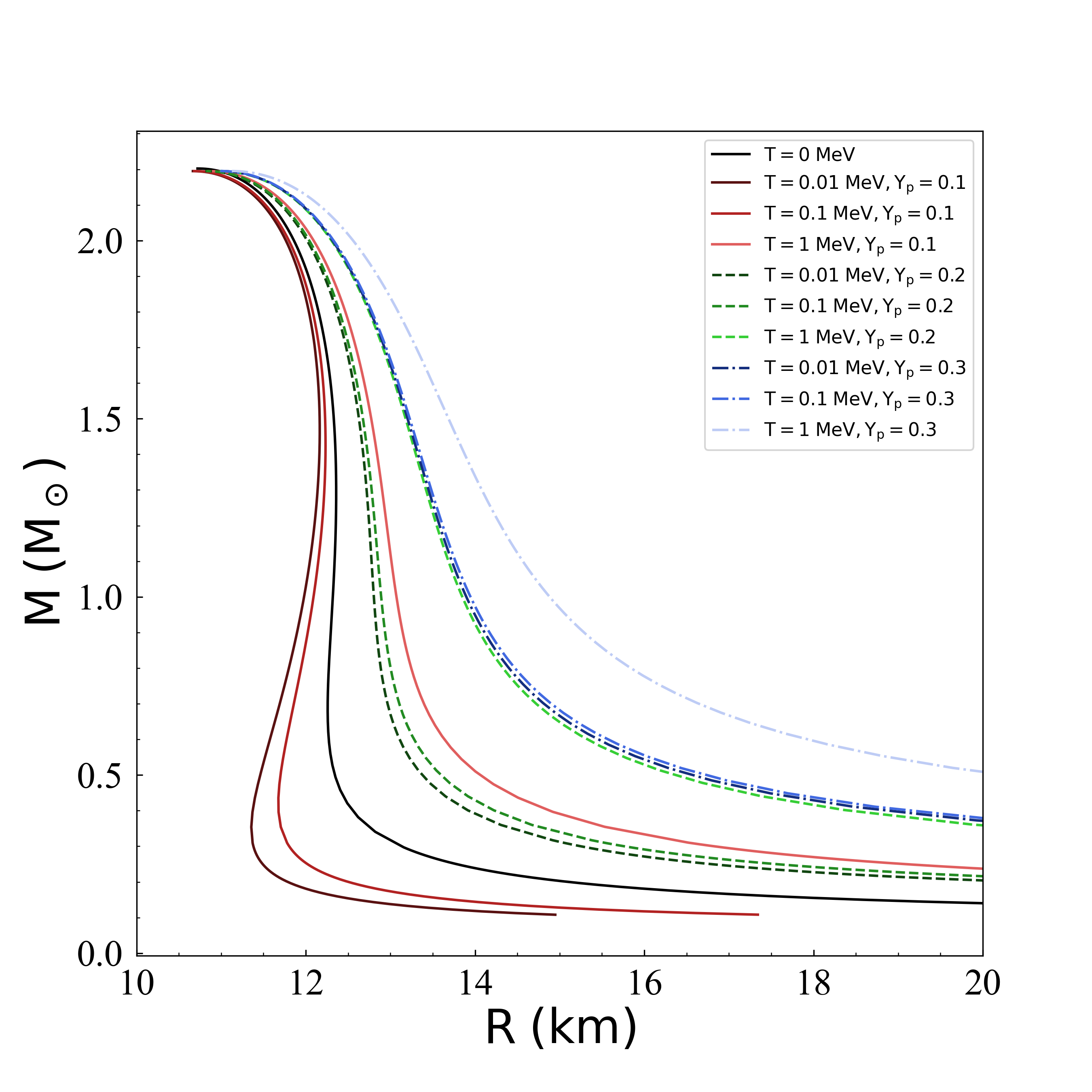}
\caption{Mass-radius dependence for (a) the four different nuclear models for various values of temperature and (b) various values of temperature and proton fraction for the MDI-APR equation of state.}
\label{MRall}
\end{figure*}

\begin{figure*}
\includegraphics[width=\textwidth]{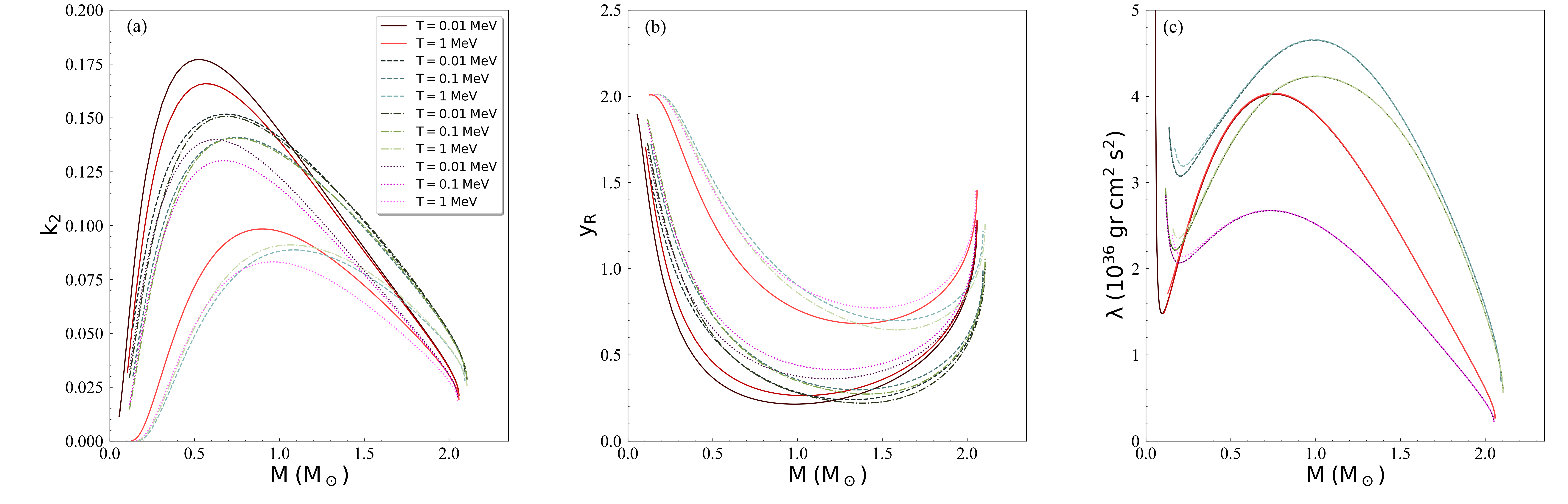}
\caption{Thermal effects on the Love number $k_2$ and individual tidal deformability.}
\label{tidalall}
\end{figure*}

\begin{figure*}
\includegraphics[width=\textwidth]{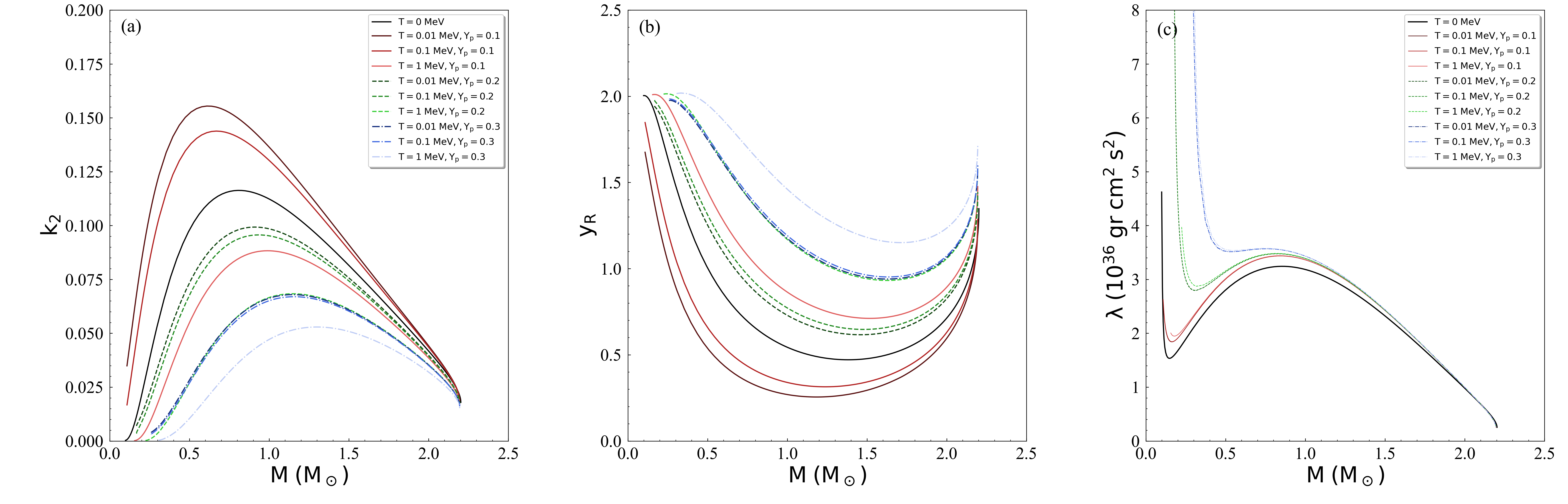}
\caption{Thermal and proton fraction effects on tidal parameters.}
\label{tidalproton}
\end{figure*}

In Figure~\ref{MRall} we display the mass-radius dependence for (a) the four different nuclear models and various values of temperature and (b) the MDI-APR equation of state for various values of temperature and proton fraction. In both cases the effect of the temperature is to increase the values of the radius (for a fixed value of mass). This effect is more pronounced especially for very high values of temperature. Moreover, the increase of the proton fraction on the warm crust leads to the increase of the size of the neutron star, enhancing the effect of temperature. 

\begin{figure*}
\includegraphics[width=\textwidth]{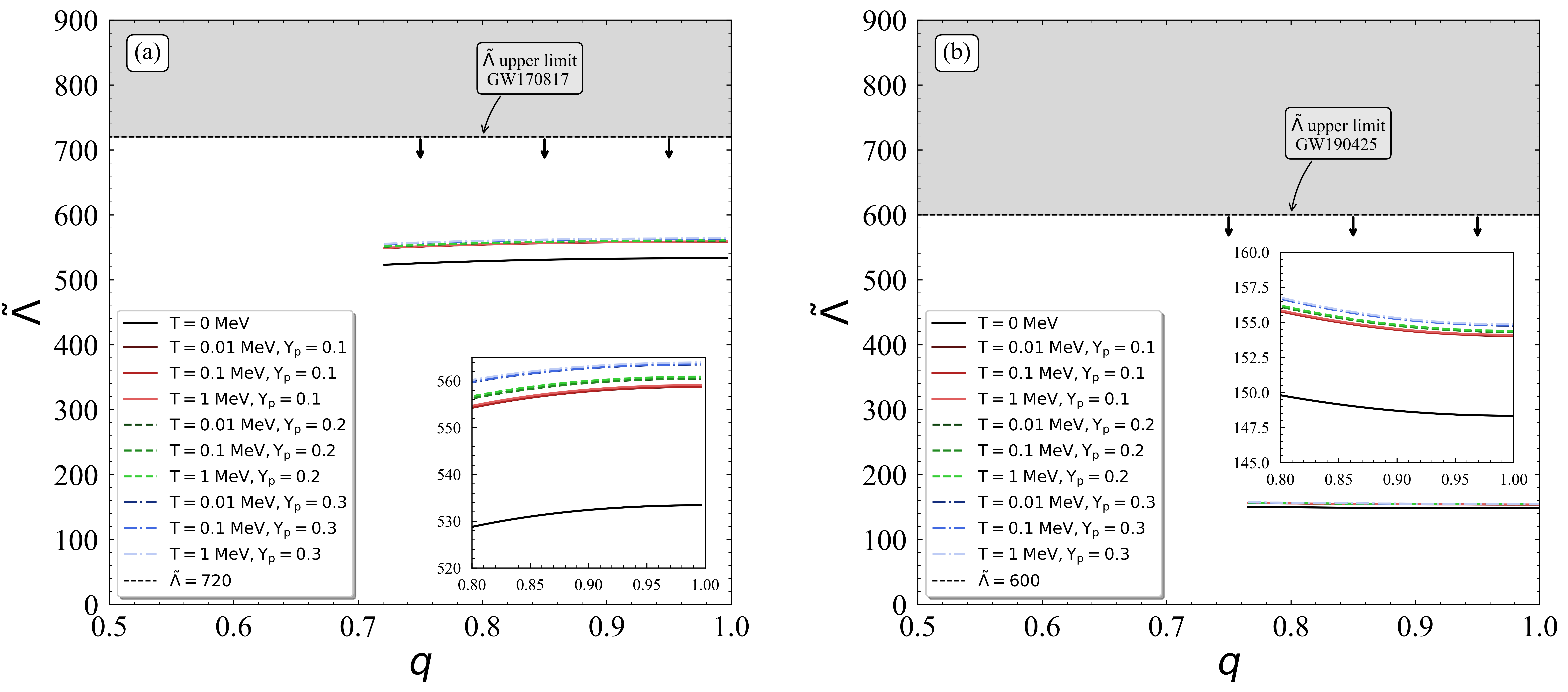}
\caption{Thermal effects on the effective tidal deformability as a function of the mass asymmetry $q=m_2/m_1$.}
\label{Lq}
\end{figure*}

Moreover, in Figure~\ref{tidalall} and Figure~\ref{tidalproton} we display the effects of temperature and proton fraction on the basic ingredients, $k_2$  and $y_R$, of the tidal deformability as well as also on $\lambda$. From Figure~\ref{tidalall}, it can be easily seen that, for any  specific equation of state,  the increase of temperature leads to a decrease in $k_2$, while the effect on the parameter $y_R$, which is consistent with the structure of the star, is also apparent. In particular, the parameter $y_R$ depends mainly on the neutron star structure, and the increment of the temperature shifts the curves to higher values (reverse behavior compared to $k_2$). However, in Figure~\ref{tidalall}(c) is displayed the surprising result that, although the radius and the Love number $k_2$ are sensitive on temperature, the corresponding tidal deformability $\lambda$ is not, both for low and high values of mass (especially close to $1.4 M_{\odot}$ which is related with the observations). Clearly there is no theoretical evidence for this behavior.

Table~\ref{temp-sigma} shows the specific values for the radius, tidal Love number $k_2$ and tidal deformability $\lambda$ for a $1.4\;M_\odot$ neutron star, related to the temperature, considering the Lattimer and Swesty~\cite{Lattimer-91} equations of state (similar results emerge for all the equations of state under consideration). Despite the decrement of $k_2$ and the increment of $R$ as the temperature increases, $\lambda$ remains almost constant in the region $T\in[0.01,1]\;\rm{MeV}$.

Figure~\ref{tidalproton}, confirms the previous finding. In particular, we studied the effect of the proton fraction on  $\lambda$  and we found that it is very small and concerns mainly the very low masses $(M\lesssim 1~M_{\odot})$ which are, after all, of no interest. 

Figure~\ref{Lq} shows the effect of the temperature and proton fraction on the effective tidal deformability $\tilde{\Lambda}$ for the MDI-APR equation of state, by using the recent observations of binary neutron star mergers~\cite{Abbott-gw170817,Abbott-gw190425}. Figure~\ref{Lq}(a) corresponds to the GW170817 event while Figure~\ref{Lq}(b) corresponds to the more recent GW190425 event. As one can observe, the effect of the temperature is present mainly on the first event with the lower component masses. Therefore, binary neutron star mergers with low value of chirp mass $\mathcal{M}_c$ could be more suitable for this kind of study.

\begin{figure}
\includegraphics[width=0.45\textwidth]{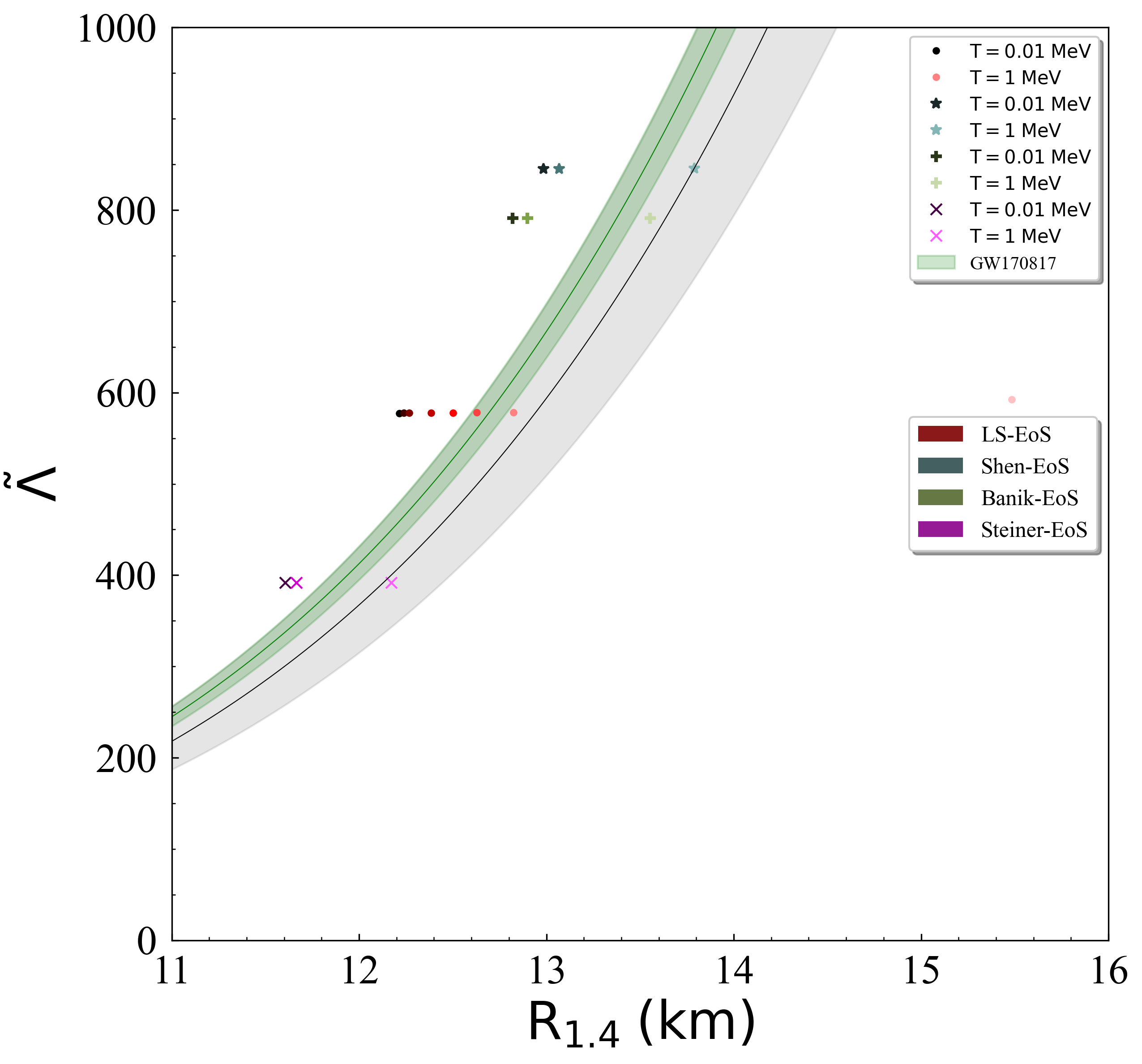}
\caption{$\tilde{\Lambda}$ versus radius $R_{1.4}$ for various equations of state.}
\label{R14}
\end{figure}

Figure~\ref{R14} demonstrates the relation between the effective tidal deformability $\tilde{\Lambda}$ and the radius $R_{1.4}$ of a $1.4\;M_\odot$ neutron star. As a reference we used the range of the component masses of the GW170817 event~\cite{Abbott-gw170817}. The shaded regions correspond to approximate relations proposed in Ref.~\cite{Zhao}. We observed that as the temperature increases, the marks are shifted to higher radii. Therefore, if we take into consideration the thermal effects, possible constraints on the radius could lead to constraints on the value of the possibly existing temperature.

\begin{table}
\begin{ruledtabular}
\begin{tabular}{lccr}
$T$ (MeV)       &   $k_2$       &    $R$ (km)   &  $\lambda$ (10$^{36}$gr cm$^2$ s$^2$) \\
\toprule
0.01    &   0.1005       &    12.21  &     2.729          \\
0.1   &   0.0984      &   12.26    &  2.730            \\
1   &   0.0788      &   12.82    &  2.730             \\
5   &   0.0315      &   15.48    &  2.790             \\
10   &   0.0127      &   18.92    &  3.020             \\
\end{tabular}
\end{ruledtabular}
\caption{Values of radius and tidal parameters related to the temperature, for a $1.4\;M_\odot$ neutron star and for the Lattimer and Swesty equations of state~\cite{Lattimer-91}.}
\label{temp-sigma}
\end{table}

\section{Conclusions} 
In the majority of the recent and previous studies, the tidal deformability is calculated considering the cold equation of state for the description of the interior of the neutron stars before the merger. According to our knowledge, this is the  first effort  dedicated to studying of thermal effects on tidal deformability (and consequently on the gravitation waves signal) just before the merger. Although there are some theoretical arguments supporting the idea of cold neutron stars before the merger (temperature less than $0.01$ MeV), other studies claim that it is possible both stars to warm enough, with their interior reaching temperatures even a few MeV. In any case, it is worth studying and also compare the two possibilities in order to gain useful insight. Moreover, it is useful to enrich our knowledge on the effect of temperature on the tidal deformability since this quantity is related mainly to the detection and analysis of gravitational waves.    

In this effort we used various sets of hot equations of state predicted in previous studies, as well as a hot equation of state suitable to describe the interior of a neutron star even for very high temperatures. We found the unexpected result, at least in terms of some solid theoretical proof, that although thermal effects, for low values of $T$ ($T<1$ MeV) are important for  $k_2$ and $R$, that is not the case for $\lambda \sim k_2 R^5$ (or $\Lambda$). The present study leads us to the conclusion that the above rule applies regardless of the applied equation of state. If this estimation proves to be correct, then it will have consequences in the way we draw conclusions from the observation of gravitational waves. Firstly, we are not able to constrain with high accuracy the radius of a neutron star with mass close to $1.4 M_{\odot}$ due to thermal uncertainties.  On the other hand, the accurate measurement of the radius, with some other reliable method can give information about the temperature of the stars in the phase before the merger (see discussion of Figure~\ref{R14}).

Concluding, we expect future observations of gravitational waves of neutron stars' merger events may shed light on this problem. Mainly, stringent constraints on $\tilde{\Lambda}-R$ dependence may help to clarify the extent of the thermal effects on the coalescing neutron stars and vice-versa.
It is among our immediate intentions to study more systematically the effect of temperature on tidal deformability using a broader set of hot equations of state. In addition, we plan to study in detail the role of the hot crust, whose structure and composition seems to be more sensitive to the effect of the temperature, due to its lower values of density (compared to the core). The above work is in progress.

\section*{Appendix}
We will try here to shed light on the insensitivity of $\lambda$ to the temperature. In general, $\lambda\sim R^5 k_2$ holds. The expansion of Love number $k_2$ up to the second order on the compactness parameter $\beta$ reads as~\cite{Piekarewicz-019}
\begin{eqnarray}
k_2&=& -\frac{1}{2}\frac{(y_R-2)}{(y_R+3)}+\frac{5}{2}\frac{(y_R^2+2y_R-6)}{(y_R+3)^2}\beta \nonumber\\
&-& \frac{5}{14}\frac{(11y_R^3+66y_R^2+52y_R-204)}{(y_R+3)^3}\beta^2+{\cal O}(\beta^3).
\end{eqnarray}
Actually, the expansion is accurate for neutron stars, even keeping only the first terms. Obviously, there is a   complicated dependence of $k_2$ on $R$ (via $\beta$) and $y_R$. This  makes the dependence of the product $k_2 R^5$ on the temperature quite complex and has no obvious explanation for its independence from the temperature (at least for low temperatures). However, our numerical calculations confirm the insensitive behavior of $\lambda$ on the temperature, at least for low temperatures (see also Table~\ref{temp-sigma}). 

\section*{Acknowledgments}
We would like to thank Prof. P. Meszaros and Prof. D. Radice for their useful corresponds and suggestions. The authors would like to thank also the Bulgarian National Science Fund (BNSF) for its support under contract KP-06-N48/1.

\end{document}